\documentclass[12pt]{article}
\usepackage{a4wide}
\usepackage{xcolor}
\usepackage{amssymb}
\usepackage{amsmath}
\begin{document}
{\renewcommand{\thefootnote}{\fnsymbol{footnote}}
\begin{center}
{\LARGE  Covariance in models of loop quantum gravity:\\ Gowdy systems}\\
\vspace{1.5em}
Martin Bojowald\footnote{e-mail address: {\tt bojowald@gravity.psu.edu}}
and Suddhasattwa Brahma\footnote{e-mail address: {\tt sxb1012@psu.edu}}\\
\vspace{0.5em}
Institute for Gravitation and the Cosmos,\\
The Pennsylvania State
University,\\
104 Davey Lab, University Park, PA 16802, USA\\
\vspace{1.5em}
\end{center}
}

\setcounter{footnote}{0}

\begin{abstract}
 Recent results in the construction of anomaly-free models of loop quantum
 gravity have shown obstacles when local physical degrees
 of freedom are present. Here, a set of no-go properties is derived in
 polarized Gowdy models, raising the question whether these systems can be
 covariant beyond a background treatment. As a side product, it is shown that
 normal deformations in classical polarized Gowdy models can be Abelianized.
\end{abstract}

\section{Introduction}

Covariance is important in cosmological models because it controls the form of
partial differential equations for inhomogeneous modes and ensures consistency
of the coupled set of equations for a smaller number of free fields. It
considerably restricts possible choices of underlying theories, for instance
of the dynamics of matter ingredients or higher-derivative corrections to
Einstein's equation.

The latter are expected also in effective equations of canonical quantum
gravity, but in such approaches covariance is not manifest. In proposed models
of loop quantum gravity, as one class of rather widely studied examples, it is
then not always clear whether covariance is realized, and to what detriment
covariance might unwittingly be broken. The main example of potentially
covariance-breaking effects is the replacement of connection components in
Hamiltonians by holonomies, a widely studied procedure which captures one of
the key ingredients of loop quantizations and gives rise to postulated
physical implications such as bounded energy densities. In \cite{SphSymmCov},
a systematic analysis of covariance in spherically symmetric or black-hole
models with modifications from loop quantum gravity has been started. Partial
no-go results have been obtained for covariant holonomy-modified models with
local matter degrees of freedom, and to date no such model is known to
exist.

Here, we extend the same methods and results to polarized Gowdy models. Also
in this context, partial no-go results will be obtained, of a form which
resembles those found in spherically symmetric models and can therefore be
taken as a sign of genericness. There seem to be obstacles to an
implementation of covariant holonomy-modified models with local degrees of
freedom, from matter or gravity. In a background treatment, local degrees of
freedom can be coupled as inhomogeneous modes to a holonomy-modified
homogeneous model. However, irrespective of whether back-reaction on the
homogeneous background is included, non-trivial covariance conditions are
present but have not been analyzed yet in existing constructions. We will
comment on hybrid models \cite{Hybrid,Hybrid2,Hybrid3} as one example. Our
statements are about holonomy-modified models characteristic of loop quantum
cosmology. They do not apply to Wheler--DeWitt type quantizations of Gowdy
models as considered for instance in
\cite{MisnerGowdy,GowdyQuadratic,BergerGowdy,HusainGowdySing,HusainSmolin,GowdyQuant,Probe2,GowdyTime,GowdyDyn,GowdyDynSchroed}.

Covariance cannot be seen in homogeneous models, the traditional setting of
loop quantum cosmology \cite{LivRev,ROPP}. At the level of effective equations,
there are only ordinary differential equations which are not subject to
additional consistency conditions from covariance. And also an equation for a
wave function, although it may be a partial differential or difference
equation, requires no such restrictions. Dynamical equations of homogeneous
cosmological models can therefore be modified at will by any putative quantum
effects, but not all versions can be minisuperspace reductions of covariant
inhomogeneous models (or of a covariant full theory of modified or quantum
gravity).

In this paper we consider polarized Gowdy systems \cite{Gowdy} as a class of
models with 1-dimensional spatial inhomogeneity and applications to
cosmology. As in \cite{SphSymmCov}, the canonical definition of covariance we
use for modified theories is based on the general form of this condition in
classical models: Instead of considering transformations generated by Lie
derivatives along space-time vector fields, one has such derivatives only for
vector fields $M^a$ tangential to spatial hypersurfaces used for the canonical
decomposition of fields. These spatial diffeomorphisms, acting on phase-space
variables, are generated by the diffeomorphism constraint $D[N^a]$. For the
remaining transformations it is sufficient to have a generator of normal
deformations of spatial hypersurfaces, given by the Hamiltonian constraint
$H[N]$, the spatial function $N$ determining the extent $N n^a$ of the
deformation along the normal vector field $n^a$. These generators have Poisson
brackets
\begin{eqnarray}
 \{D[M_1^a],D[M_2^a]\} &=& D[{\cal L}_{M_1}M_2^a] \label{DD}\\
 \{H[N],D[M^a]\} &=& -H[{\cal L}_MN] \label{HD}\\
 \{H[N_1],H[N_2]\} &=& D[q^{ab}(N_1\partial_bN_2-N_2\partial_bN_1)] \label{HH}
\end{eqnarray}
with structure functions in the last line, given by the inverse spatial metric
$q^{ab}$ \cite{DiracHamGR,ADM}.

A modified or quantized canonical theory must have at least a classical limit
in which (\ref{DD})--(\ref{HH}) are realized. For non-classical solutions, the
brackets may be subject to quantum corrections but must still close for an
anomaly-free theory: Since the constraints generate gauge transformations,
there must be analogs of the classical constraints $D[M^a]$ and $H[N]$ with
brackets which are closed under all circumstances (not just in the classical
limit). As discussed in \cite{SphSymmCov}, there are therefore two conditions
for a covariant theory: (i) Anomaly-free gauge generators and (ii) a classical
limit in which the hypersurface-deformation brackets (\ref{DD})--(\ref{HH})
are obtained. As shown in \cite{SphSymmCov}, building on results of
\cite{LoopSchwarz2}, condition (ii) is not necessarily a consequence of
condition (i).

An important part of the conditions for covariance is that they refer to the
off-shell brackets when the constraints are not necessarily zero. This feature
is analogous to the usual space-time definition of a covariant theory as one
with a Lagrangian covariant under tensor transformations. The conceptual
reason for the prominence of off-shell structures is the classical picture of
space-time as a background on which different kinds of matter fields can be
put. Even if back-reaction is included and one does not restrict equations to
those for a field on a fixed background, one thinks of space-time as an
independent ingredient which is covariant in its own right, irrespective of
the matter fields coupled to it. After all, the Einstein tensor of any
space-time, independently of solutions to field equations or the inclusion of
back-reaction, obeys the contracted Bianchi identity, which in canonical form
is equivalent to a version of (\ref{DD})--(\ref{HH}) \cite{KucharHypI}. For a
consistent matter coupling, one therefore requires the local conservation law
for the matter stress-energy tensor, again independently of solutions to field
equations. Also the local conservation law is equivalent to a version of
(\ref{DD})--(\ref{HH}) for matter Hamiltonians \cite{Energy}. In both cases,
the form of off-shell brackets is crucial, which we will analyze for modified
Gowdy models in the present paper.

\section{Modified theories with local degrees of freedom?}

Since the algebraic structure of modified Gowdy models is closely related to
the one of spherically symmetric models discussed in \cite{SphSymmCov}, we
will begin with a brief review of these existing results.

\subsection{Spherical symmetry}

Using triad variables $E^x$ and $E^{\varphi}$ with canonically conjugate
extrinsic-curvature components $K_x$ and $K_{\varphi}$, the gravitational
contribution to the spherically symmetric Hamiltonian constraint is
\begin{equation}
 H[N]=-\frac{1}{2G}\int{\rm d}x N(x) \left(|E^x|^{-\frac{1}{2}}
   E^{\varphi}K_{\varphi}^2+
2 |E^x|^{\frac{1}{2}} K_{\varphi}K_x
+ |E^x|^{-\frac{1}{2}}(1-\Gamma_{\varphi}^2)E^{\varphi}+
2\Gamma_{\varphi}'|E^x|^\frac{1}{2}\right)
\end{equation}
where $\Gamma_{\varphi}=-(E^x)'/2E^{\varphi}$ (see
\cite{SphSymm,SphSymmHam}). If one adds to this the matter Hamiltonian,
for instance
\begin{equation}
 H_{\phi}[N] = \frac{1}{8G} \int{\rm d}x N(x)
     \frac{1}{\sqrt{|E^x|}E^{\varphi}} \left(P_{\phi}^2+4(E^x)^2
       (\phi')^2\right)
\end{equation}
for a scalar field $\phi$ with momentum $P_{\phi}$, the
hypersurface-deformation brackets are realized in combination with the
diffeomorphism constraint
 \begin{equation}
 D[M] = \frac{1}{G} \int{\rm d}x M(x) \left(-\frac{1}{2}(E^x)'K_x+K_{\varphi}'
   E^{\varphi} +GP_{\phi}\phi'\right)\,.
\end{equation}
Instead of the full spatial metric $q^{ab}$, the structure functions are given
by the radial component $|E^x|/(E^{\varphi})^2$ of a spherically
symmetric inverse spatial metric.

In order to eliminate the structure functions, \cite{LoopSchwarz2} introduced
a linear combination of the constraints so that the normal part of
hypersurface deformations is replaced by an Abelian bracket. In this process,
$H[N]$ is replaced by a new constraint
\begin{eqnarray}\label{MattHam}
 C[N]&=& \int{\rm d}x N(x) \Biggl(-\frac{1}{2} \frac{(E^x)'}{\sqrt{|E^x|}}
     (1+K_{\varphi}^2)- 2\sqrt{|E^x|} K_{\varphi}K_{\varphi}'\\
 &&+
     \frac{(E^x)'}{8\sqrt{|E^x|}(E^{\varphi})^2} \left(4E^x(E^x)''+
       ((E^x)')^2\right) - \frac{1}{2} \frac{((E^x)')^2 \sqrt{|E^x|}
       (E^{\varphi})'}{(E^{\varphi})^3}\nonumber\\
&& + \frac{1}{8}
     \frac{(E^x)'}{\sqrt{|E^x|}(E^{\varphi})^2} \left(P_{\phi}^2+4(E^x)^2
       (\phi')^2\right)- \sqrt{|E^x|} \frac{K_{\varphi}}{E^{\varphi}}
     P_{\phi}\phi'\Biggr)\,. \nonumber
\end{eqnarray}
While the pair $(C[N],D[M])$ does not obey the hypersurface-deformation
brackets (\ref{DD})--(\ref{HH}), it has a reduced phase space equivalent to
the one of the original system. Quantizing the partially Abelianized system
should be easier, as proposed in \cite{LoopSchwarz2} in combination with a
background treatment.

As part of the loop quantization performed in \cite{LoopSchwarz2}, one
modifies the dependence of (\ref{MattHam}) on $K_{\varphi}$ by replacing it
with some bounded function $f(K_{\varphi})$ in order to model the appearance
of holonomies in loop quantum gravity. However, as there are three different
terms in (\ref{MattHam}) depending on $K_{\varphi}$, there could in general be
three replacement functions which need not be equal but should be related in
some way for a consistent theory in which the brackets still close and
implement covariance. In \cite{LoopSchwarz2}, this question has been
circumvented by the background treatment in which one first considers only the
gravitational part of $C[N]$, which happens to be a total derivative. Upon
integrating by parts, there is only one term depending on $K_{\varphi}$, which
can easily be modified by a single function $f(K_{\varphi})$ while keeping the
constraint bracket Abelian.

However, as shown in \cite{SphSymmCov}, the modification is consistent with
covariance only if the different $K_{\varphi}$-dependent terms in the original
constraint are modified in strictly related ways, of a form equivalent to what
had been found earlier by effective methods \cite{JR,HigherSpatial}: The
gravitational part of the modified Hamiltonian constraint then has to be of
the form
\begin{eqnarray}\label{modHam}
 H[N]&=&-\frac{1}{2G}\int{\rm d}x N(x) \left(|E^x|^{-\frac{1}{2}}
   E^{\varphi}f_1\left(K_{\varphi}\right)+
2 |E^x|^{\frac{1}{2}} f_2\left(K_{\varphi}\right)K_x\right.\\
&&+ \left.|E^x|^{-\frac{1}{2}}(1-\Gamma_{\varphi}^2)E^{\varphi}+
2\Gamma_{\varphi}'|E^x|^\frac{1}{2}\right) \nonumber
\end{eqnarray}
with
\begin{equation} \label{f2f1}
 2f_2=\frac{{\rm  d}f_1}{{\rm d}K_{\varphi}}\,.
\end{equation}
As a consequence, the hypersurface-deformation brackets are modified at large
curvature and show signature change \cite{Action,SigChange,SigImpl}: The
classical structure function is multiplied with
\begin{equation}\label{betaSph}
 \beta = \frac{{\rm d}f_2}{{\rm d}K_{\varphi}}= \frac{1}{2}\frac{{\rm
     d}^2f_1}{{\rm d}K_{\varphi}^2}
\end{equation}
which is negative around a local maximum of the modification function
$f_1(K_{\varphi})$.

Moreover, while the classical system is still Abelian in the presence of a
non-zero matter Hamiltonian, no consistent modification has been found. It is
therefore unclear whether modified combined systems of gravity and matter can
be covariant. We now turn to Gowdy models in order to test whether the problem
rests with the form of matter terms or is implied by the general presence of
local degrees of freedom.

\subsection{Polarized Gowdy models}

In contrast to spherically symmetric models, polarized Gowdy models have local
physical degrees of freedom even if there is no matter. At the kinematical
level, on which off-shell questions about constraints are addressed, the local
degree of freedom is included by an additional canonical pair of
fields. Nevertheless, the structure of the constraints and their algebraic
properties are closely related to those of spherically symmetric models, so
that a comparison can easily be done and is quite instructive.

\subsubsection{Variables}

In Gowdy models, the inhomogeneous coordinate is traditionally called
$\theta$, while $x$, used in spherically symmetric models for the radial
coordinate, is part of a pair $(x,y)$ of coordinates along two independent
homogeneous directions. In a real connection formulation
\cite{EinsteinRosenAsh} (see \cite{HusainSmolin} for complex variables), there
are three triad variables $(\epsilon, E^x, E^y)$ and canonical momenta $({\cal
  A},K_x,K_y)$. They appear in the diffeomorphism constraint in standard form
\begin{eqnarray}\label{Diffeo1}
D[N^\theta]=\frac{1}{8\pi G}\int\mathrm{d}\theta
N^\theta(\theta)\left(K_x'E^x+K_y'E^y- \varepsilon'\mathcal{A}  \right)
\end{eqnarray}
while the Hamiltonian constraint is
\begin{eqnarray}\label{Ham1}
 H[N]&&=\frac{-1}{8\pi G}\int\mathrm{d}\theta N(\theta)
\Bigg[f(K_x, K_y)(E^x)^{1/2}(E^y)^{1/2}\varepsilon^{-1/2} +
g_1(K_x, K_y)\mathcal{A}(E^x)^{1/2}(E^y)^{-1/2}\varepsilon^{1/2}\nonumber\\
&&+ g_2(K_x, K_y)\mathcal{A}(E^x)^{-1/2}(E^y)^{1/2}\varepsilon^{1/2}-
\frac{1}{4}(E^x)^{-1/2}(E^y)^{-1/2}\varepsilon^{-1/2}(\varepsilon')^{2}\nonumber\\
&&-\frac{1}{4}(E^x)^{-1/2}(E^y)^{-5/2}\varepsilon^{3/2}(E^{y\prime})^{2}-
\frac{1}{4}(E^x)^{-5/2}(E^y)^{-1/2}\varepsilon^{3/2}(E^{x\prime})^{2}\nonumber\\
&&+\frac{1}{2}(E^x)^{-3/2}(E^y)^{-3/2}\varepsilon^{3/2}(E^{x\prime})(E^{y\prime})
+\frac{1}{2}(E^x)^{-3/2}(E^y)^{-1/2}\varepsilon^{1/2}(E^{x\prime})
\varepsilon'\nonumber\\
&&+\frac{1}{2}(E^x)^{-1/2}(E^y)^{-3/2}\varepsilon^{1/2}(E^{y\prime})
\varepsilon'-(E^x)^{-1/2}(E^y)^{-1/2}\varepsilon^{1/2}\varepsilon''\Bigg]\,.
\end{eqnarray}
Classically, $f(K_x, K_y)=K_xK_y$, $g_1(K_x, K_y)=K_x$ and $g_2(K_x, K_y)=K_y$
but as before, the dependence may be modified based on quantum-geometry
effects such as the use of holonomies in loop quantum gravity. The classical
structure function in the bracket of two normal deformations is
$\epsilon^2/E^xE^y$.

\subsubsection{Structure of modification functions}
\label{s:Structure}

The question of consistent deformations of the classical brackets can be split
in two: (a) What are the conditions on modification functions $f$, $g_1$ and
$g_2$ for the brackets to be closed? And (b), what are the possible
modifications of the classical structure function? In order to address (b),
(a) must be solved since meaningful structure functions require a consistent
set of brackets. However, at a purely formal level one may analyze (b) without
first solving (a), in order to study possible features of interest in
deformations of the brackets. The main effect seen in this way is signature
change \cite{Action,SigChange,SigImpl}, given by a change of sign of the
structure function, which would always be positive in a classical Lorentzian
theory. In the first part of this subsection, we analyze (b) for Gowdy models,
postponing detailed derivations of Poisson brackets to the subsequent
consideration of (a).

\paragraph{Deformations and the ubiquity of signature change.}

From the relations to be presented soon, it follows that an anomaly-free
modification of the Hamiltonian constraint (\ref{Ham1}) requires the following
equation to hold for all values of the canonical fields: We must have
\begin{eqnarray}\label{termscancel}
&&\left[\frac{1}{2}(E^y)^{-2}\varepsilon E^{y\prime} -
  \frac{1}{2}(E^y)^{-1}\varepsilon' -
  \frac{1}{2}(E^x)^{-1}(E^y)^{-1}\varepsilon
  E^{x\prime}\right](f_{,K_y}-g_1)\nonumber\\
&&+\left[\frac{1}{2}(E^x)^{-2}\varepsilon E^{x\prime} -
  \frac{1}{2}(E^x)^{-1}\varepsilon' -
  \frac{1}{2}(E^x)^{-1}(E^y)^{-1}\varepsilon
  E^{y\prime}\right](f_{,K_x}-g_2)\nonumber\\
&&+\left[\frac{1}{2}\mathcal{A}(E^x)^{-2}(E^y)^{-1}\varepsilon^2 E^{x\prime}
  -\frac{1}{2}\mathcal{A}(E^x)^{-1}(E^y)^{-2}\varepsilon^2
  E^{y\prime}\right](g_{1,K_x}-g_{2,K_y})\nonumber\\
&&+\frac{1}{2}\mathcal{A}\varepsilon^2\left[\frac{E^{x\prime}}{E^x}
  -\frac{E^{y\prime}}{E^y}\right]\left[\frac{g_{2,K_x}}{E_x^2}-\frac{g_{1,K_y}}{E_y^2}
\right]=0
\end{eqnarray}
for all terms in the $\{H,H\}$-bracket that cannot contribute to a
diffeomorphism constraint to cancel out.  (As usual, commas in subscripts
indicate partial derivatives by the appended variable(s).)  All lines must
vanish individually since their coefficients are composed of different
functions of the canonical variables and their derivatives. (Otherwise,
additional constraints on the phase-space variables would be imposed.)
Requiring the first two lines in (\ref{termscancel}) to be zero gives two
conditions,
\begin{eqnarray}\label{conds1}
&&g_1(K_x,K_y)=\frac{\partial f(K_x,K_y)}{\partial K_y}\nonumber\\
&&g_2(K_x,K_y)=\frac{\partial f(K_x,K_y)}{\partial K_x}\,,
\end{eqnarray}
for two of the three free modification functions.  These conditions
automatically make the third line in (\ref{termscancel}) vanish, owing to the
equality of mixed partial derivatives. The last line in (\ref{termscancel}) is
zero if and only if
\begin{eqnarray}\label{conds2}
\frac{\partial^2 f}{\partial K_x^2}=\frac{1}{(E^x/E^y)^2} \frac{\partial^2
  f}{\partial K_y^2}\,,
\end{eqnarray}
providing some kind of wave equation for the remaining modification function.

At this stage, we note the first important difference to the spherically
symmetric case where it is possible to have modification functions depending
only on the curvature variables. In polarized Gowdy models, by contrast, any
function $f$, which solves condition (\ref{conds2}) and differs from the
classical limit by cubic or higher-order terms in curvature variables, must
also depend on some of the triad variables. (Spherically symmetric models can
be seen as reductions of polarized Gowdy models with $E^x=E^y$, so that the
triad dependence disappears from (\ref{conds2}).) We must therefore go back
and rederive brackets of (\ref{Ham1}), because in (\ref{termscancel}) we have
assumed that $f$ depends only on curvature variables.

However, we may proceed further without rederiving the more-complete brackets,
addressing question (b) without solving problem (a) introduced in the
beginning discussion of this subsection.  If (\ref{termscancel}) is assumed to
hold, the remaining terms of the $\{H,H\}$-bracket, containing
  factors that appear in the diffeomorphism constraint, are
\begin{eqnarray}\label{remterms1}
&&-\frac{1}{8\pi G}\int\mathrm{d}\theta (MN'-NM')
\frac{\varepsilon^2\mathcal{A}\varepsilon'}{E^xE^y}
\left[\frac{\partial^2f}{\partial K_x\partial K_y} +\frac{1}{2}
  \frac{\partial^2f}{\partial K_y^2}\frac{E^x}{E^y}
  +\frac{1}{2} \frac{\partial^2f}{\partial
      K_x^2}\frac{E^y}{E^x}\right]\\
&&+ \frac{1}{8\pi G}\int\mathrm{d}\theta (MN'-NM')
\frac{\varepsilon^2}{E^xE^y}\Bigg[\frac{\partial^2f}{\partial
    K_x\partial K_y}(K_x'E^x+K_y'E^y) +
\frac{\partial^2f}{\partial
    K_y^2}E^xK_y'+\frac{\partial^2f}{\partial
    K_x^2}E^yK_x'\Bigg]\,,\nonumber
\end{eqnarray}
where we have already used condition (\ref{conds1}) to simplify the terms. If
we insert (\ref{conds2}) in (\ref{remterms1}), we can simplify the structure
function in front of terms contributing to the diffeomorphism constraint. The
resulting expression is
\begin{eqnarray}\label{remterms2}
\frac{1}{8\pi G}\int\mathrm{d}\theta (MN'-NM')
\frac{\varepsilon^2}{E^xE^y}\left[\frac{\partial^2f}{\partial
      K_x\partial K_y} + \frac{\partial^2f}{\partial
      K_y^2}\frac{E^x}{E^y}\right]\left[K_x'E^x+K_y'E^y
  -\mathcal{A}\varepsilon' \right]
\end{eqnarray}
where, in addition to the classical structure function
$\varepsilon^2/(E^xE^y)$, we have a deformation function
\begin{equation} \label{beta}
 \beta=\frac{\partial^2f}{\partial K_x\partial K_y} +
\frac{\partial^2f}{\partial
    K_y^2}\frac{E^x}{E^y}\,.
\end{equation}

Although this function is more complicated than its spherically symmetric
analog (\ref{betaSph}), it is still possible to show that for any modification
function $f$ with a local maximum, the modified structure function has
negative values, $\beta<0$.  In order to do so, we solve (\ref{conds2}) by
requiring $f$ to have the form
$f(K_x,K_y,E^x,E^y)=f_1(E^xK_x+E^yK_y)+f_2(E^xK_x-E^yK_y)$ with two free
functions $f_1$ and $f_2$ of one variable. The positions of local maxima of
$f$ are determined by properties of the following derivatives:
\begin{eqnarray}\label{sigchange1}
f_{,K_xK_x}&=&(E^x)^2\left[\ddot{f}_1+\ddot{f}_2\right]\nonumber\\
f_{,K_yK_y}&=&(E^y)^2\left[\ddot{f}_1+\ddot{f}_2\right]\nonumber\\
f_{,K_xK_y}&=&E^xE^y\left[\ddot{f}_1-\ddot{f}_2\right]\,,
\end{eqnarray}
where a dot over a function denotes a derivative with respect to its
argument. At a local maximum, the standard conditions $f_{,K_xK_x}<0$ and
$f_{,K_xK_x}f_{,K_yK_y}-(f_{,K_xK_y})^2>0$ imply
\begin{equation}\label{sigchange2}
\ddot{f}_1+\ddot{f}_2<0 \quad\text{and}\quad
\ddot{f}_1\ddot{f}_2>0\,.
\end{equation}
Therefore, both $\ddot{f}_1$ and $\ddot{f}_2$ have to be negative.

The deformation function $\beta$ in (\ref{beta}) is proportional to the first
of these expressions,
\begin{eqnarray}\label{sigchange3}
\beta=2E^xE^y\ddot{f}_1,
\end{eqnarray}
so that it turns negative around a local maximum of $f$. The formal aspects of
deformation functions, disregarding full anomaly-freedom for now, is therefore
in complete agreement with previous investigations in spherically symmetric
models \cite{JR} and for cosmological perturbations \cite{ScalarHolInv}. (See
also \cite{DeformedCosmo}.) Around local maxima of modification functions, the
modified structure function in the bracket of normal hypersurface deformations
is negative, as it is for Euclidean space. Hyperbolic wave equations are then
replaced by elliptic equations which do not allow deterministic propagation
through such a region, typically at large curvature. Implications have been
studied for cosmological \cite{SigImpl} and black-hole models \cite{Loss}.

\paragraph{Closure?}

We have seen that we have to generalize the dependence of modification
functions on the canonical variables in order to solve part (a) of the
question of consistent deformations of the bracket of Hamiltonian constraints.
The class of solutions we will find has the classical dependence on curvature
variables, so that holonomy modifications are ruled out in modified models as
assumed here.

Our more-general ansatz is
\begin{eqnarray}\label{Ham2}
&& H[N]=\frac{-1}{8\pi G}\int\mathrm{d}\theta N(\theta)\Bigg[f(K_x, K_y, E^x,
E^y, \varepsilon) + g(K_x, K_y, E^x, E^y, \varepsilon)\mathcal{A}\nonumber\\
&&\hspace{2cm}-\frac{1}{4}(E^x)^{-1/2}(E^y)^{-1/2}\varepsilon^{-1/2}
(\varepsilon')^{2}
-\frac{1}{4}(E^x)^{-1/2}(E^y)^{-5/2}\varepsilon^{3/2}(E^{y\prime})^{2}\nonumber\\
&&\hspace{2cm}-\frac{1}{4}(E^x)^{-5/2}(E^y)^{-1/2}\varepsilon^{3/2}(E^{x\prime})^{2}
+\frac{1}{2}(E^x)^{-3/2}(E^y)^{-3/2}\varepsilon^{3/2}(E^{x\prime})(E^{y\prime})
\nonumber\\
&&\hspace{2cm}
+\frac{1}{2}(E^x)^{-3/2}(E^y)^{-1/2}\varepsilon^{1/2}(E^{x\prime})\varepsilon'
+\frac{1}{2}(E^x)^{-1/2}(E^y)^{-3/2}\varepsilon^{1/2}(E^{y\prime})\varepsilon'
\nonumber\\
&&\hspace{2cm} -(E^x)^{-1/2}(E^y)^{-1/2}\varepsilon^{1/2}\varepsilon''\Bigg]\,.
\end{eqnarray}
At this stage, we only assume that the modified Hamiltonian constraint is
linear in $\mathcal{A}$, motivated by the result that a non-linear dependence
on the connection component in the inhomogeneous direction is difficult to
achieve in spherically symmetric models even if a derivative expansion is
allowed for \cite{HigherSpatial}. We are therefore considering only point-wise
holonomy corrections with angular curvature or connection components, setting
aside the question of possible non-local modifications that holonomies in the
inhomogeneous direction are expected to entail.

As before, only the first and second terms give non-zero contributions to the
$\{H,H\}$-bracket. Providing more details than before, we list the integrands
of all of them, not writing the common factor of smearing functions
$(M'N-N'M)$.  The first term gives rise to
\begin{eqnarray}\label{NGtermsfrom1}
&&\frac{1}{2}f_{,K_y}(E^x)^{-1/2}(E^y)^{-5/2}\varepsilon^{3/2} E^{y\prime}
+\frac{1}{2}f_{,K_x}(E^x)^{-5/2}(E^y)^{-1/2}\varepsilon^{3/2} E^{x\prime}
\nonumber\\
&& - \frac{1}{2}f_{,K_y}(E^x)^{-3/2}(E^y)^{-3/2}\varepsilon^{3/2} E^{x\prime}
- \frac{1}{2}f_{,K_x}(E^x)^{-3/2}(E^y)^{-3/2}\varepsilon^{3/2} E^{y\prime}
\nonumber\\
&&-\frac{1}{2}f_{,K_x}(E^x)^{-3/2}(E^y)^{-1/2}\varepsilon^{1/2}\varepsilon'
-\frac{1}{2}f_{,K_y}(E^x)^{-1/2}(E^y)^{-3/2}\varepsilon^{1/2}\varepsilon'\,,
\end{eqnarray}
whereas the various commutators with the second term yield
\begin{eqnarray}\label{NGtermsfrom2}
&&\frac{1}{2}g(E^x)^{-1/2}(E^y)^{-1/2}\varepsilon^{-1/2}\varepsilon'
+\frac{1}{2}g_{,K_y}\mathcal{A}(E^x)^{-1/2}(E^y)^{-5/2}\varepsilon^{3/2}
E^{y\prime}\nonumber\\
&&+\frac{1}{2}g_{,K_x}\mathcal{A}(E^x)^{-5/2}(E^y)^{-1/2}\varepsilon^{3/2}E^{x\prime}
-\frac{1}{2}g_{,K_y}\mathcal{A}(E^x)^{-3/2}(E^y)^{-3/2}\varepsilon^{3/2}E^{x\prime}
\nonumber\\
&&-\frac{1}{2}g_{,K_x}\mathcal{A}(E^x)^{-3/2}(E^y)^{-3/2}\varepsilon^{3/2}E^{y\prime}
-\frac{1}{2}g_{,K_x}\mathcal{A}(E^x)^{-3/2}(E^y)^{-1/2}\varepsilon^{1/2}\varepsilon'
\nonumber\\
&&-\frac{1}{2}g(E^x)^{-3/2}(E^y)^{-1/2}\varepsilon^{1/2}E^{x\prime}
-\frac{1}{2}g(E^x)^{-1/2}(E^y)^{-3/2}\varepsilon^{1/2}E^{y\prime}\nonumber\\
&&-\frac{1}{2}g_{,K_y}\mathcal{A}(E^x)^{-1/2}(E^y)^{-3/2}\varepsilon^{1/2}
\varepsilon'
-\frac{1}{2}g(E^x)^{-3/2}(E^y)^{-1/2}\varepsilon^{1/2}E^{x\prime}\nonumber\\
&&-\frac{1}{2}g(E^x)^{-1/2}(E^y)^{-3/2}\varepsilon^{1/2}E^{y\prime}
+\frac{1}{2}g(E^x)^{-1/2}(E^y)^{-1/2}\varepsilon^{-1/2}\varepsilon'\nonumber\\
&&+g(E^x)^{-3/2}(E^y)^{-1/2}\varepsilon^{1/2}E^{x\prime}
+g(E^x)^{-1/2}(E^y)^{-3/2}\varepsilon^{1/2}E^{y\prime}
-g(E^x)^{-1/2}(E^y)^{-1/2}\varepsilon^{-1/2}\varepsilon'\nonumber\\
&&+\left[g_{,K_x}K_x' +g_{,K_y}K_y' +g_{,E^x}E^{x\prime} +g_{,E^y}E^{y\prime}
  +g_{,\varepsilon}\varepsilon'\right](E^x)^{-1/2}(E^y)^{-1/2}\varepsilon^{1/2}\,.
\end{eqnarray}

Several of these terms cancel each other so that the combined expression can
be simplified.  For the bracket to be proportional to the diffeomorphism
constraint, terms in (\ref{NGtermsfrom1}) and (\ref{NGtermsfrom2}) not
proportional to $\mathcal{A}\varepsilon'$, $K_x'$ or $K_y'$ must vanish:
\begin{eqnarray}\label{termscancel1}
&&E^{y\prime}\left[\frac{1}{2}f_{,K_y}(E^x)^{-1/2}(E^y)^{-5/2}\varepsilon^{3/2}
-\frac{1}{2}f_{,K_x}(E^x)^{-3/2}(E^y)^{-3/2}\varepsilon^{3/2}
+g_{,E^y}(E^x)^{-1/2}(E^y)^{-1/2}\varepsilon^{1/2}\right]\nonumber\\
&&+E^{x\prime}\left[\frac{1}{2}f_{,K_x}(E^x)^{-5/2}(E^y)^{-1/2}\varepsilon^{3/2}
-\frac{1}{2}f_{,K_y}(E^x)^{-3/2}(E^y)^{-3/2}\varepsilon^{3/2}
+g_{,E^x}(E^x)^{-1/2}(E^y)^{-1/2}\varepsilon^{1/2}\right]\nonumber\\
&&-\varepsilon'\left[\frac{1}{2}f_{,K_x}(E^x)^{-3/2}(E^y)^{-1/2}\varepsilon^{1/2}
\frac{1}{2}f_{,K_y}(E^x)^{-1/2}(E^y)^{-3/2}\varepsilon^{1/2}
+g_{,\varepsilon}(E^x)^{-1/2}(E^y)^{-1/2}\varepsilon^{1/2}\right]\nonumber\\
&&+\mathcal{A}E^{y\prime}\left[g_{,K_y}(E^x)^{-1/2}(E^y)^{-5/2}\varepsilon^{3/2}
+g_{,K_x}(E^x)^{-3/2}(E^y)^{-3/2}\varepsilon^{3/2}\right]\nonumber\\
&&+\mathcal{A}E^{x\prime}\left[g_{,K_x}(E^x)^{-5/2}(E^y)^{-1/2}\varepsilon^{3/2}
+g_{,K_y}(E^x)^{-3/2}(E^y)^{-3/2}\varepsilon^{3/2}\right]=0\,.
\end{eqnarray}
As before, all lines in (\ref{termscancel1}) must vanish individually when
they have different coefficients. We obtain four independent conditions on the
correction functions:
\begin{eqnarray}
\frac{\partial g}{\partial
    \varepsilon}&=&\frac{1}{E^x}\frac{\partial f}{\partial
    K_x} + \frac{1}{E^y} \frac{\partial f}{\partial
    K_y}\label{NGcond1}\\
\frac{\partial g}{\partial K_x}&=& \frac{1}{E^y/E^x}
\frac{\partial g}{\partial K_y}\label{NGcond2}\\
\frac{\partial g}{\partial E^y}&=&-\frac{\varepsilon}{(E^y)^2}
\frac{\partial f}{\partial K_y} +
\frac{\varepsilon}{E^xE^y} \frac{\partial f}{\partial
    K_x}\label{NGcond3}\\
\frac{\partial g}{\partial E^x}&=&-\frac{\varepsilon}{(E^x)^2}
\frac{\partial f}{\partial K_x} +
\frac{\varepsilon}{E^xE^y} \frac{\partial f}{\partial
    K_y}\label{NGcond4}\,.
\end{eqnarray}

From (\ref{NGcond2}), $g$ has to be of the form
\begin{eqnarray}\label{Sol1}
g(K_x, K_y, E^x, E^y, \varepsilon)=g_1(E^xK_x+E^yK_y) \,g_2(E^x, E^y,
\varepsilon)\,.
\end{eqnarray}
Using this form of the correction function in (\ref{NGcond3}) and
(\ref{NGcond4}), respectively, gives
\begin{eqnarray}
-\frac{1}{E^y} \frac{\partial f}{\partial K_y}
+\frac{1}{E^x}\frac{\partial f}{\partial K_x}&=&
\frac{E^y}{\varepsilon}\left[g_1\frac{\partial g_2}{\partial E^y}
  +K_yg_2\dot{g_1}\right]\label{Sol2}\\
\frac{1}{E^y} \frac{\partial f}{\partial K_y}
-\frac{1}{E^x}\frac{\partial f}{\partial K_x}&=&
\frac{E^x}{\varepsilon}\left[g_1\frac{\partial g_2}{\partial E^x}
  +K_xg_2\dot{g_1}\right]\label{Sol3}\,.
\end{eqnarray}
Combining (\ref{Sol2}) and (\ref{Sol3}),
\begin{eqnarray}\label{Sol4}
g_1\left[E^y\frac{\partial g_2}{\partial E^y}
  +E^x\frac{\partial g_2}{\partial E^x} \right]
+g_2\dot{g_1}\left[E^xK_x+E^yK_y\right]&=&0\,.
\end{eqnarray}

We can try to solve the final differential equation by employing separation of
variables. Abbreviating $\Theta := E^xK_x+E^yK_y$, we have
\begin{eqnarray}\label{Sol4.1}
\frac{1}{g_2}\left[E^y\frac{\partial g_2}{\partial E^y}
  +E^x\frac{\partial g_2}{\partial E^x}
\right]=-\frac{\Theta}{g_1}\frac{{\rm d}g_1}{{\rm d}\Theta}\,.
\end{eqnarray}
The left-hand side is a function of the triad components alone whereas the
right-hand side depends on a particular combination of triads and connection
coefficients. Thus, they must both be equal to some constant, say, $c$.  The
functions $g_1, g_2$ then satisfy the differential equations
\begin{eqnarray}\label{Sol5}
\frac{{\rm d}g_1}{g_1}&=&c\frac{{\rm d}\Theta}{\Theta}\\
E^y\frac{\partial g_2}{\partial E^y} +E^x\frac{\partial
    g_2}{\partial E^x}&=&-c g_2
\end{eqnarray}
with solutions
\begin{eqnarray}\label{Sol6}
g_1\left(E^xK_x+E^yK_y\right)&=&c_1 \left[E^xK_x+E^yK_y\right]^c\,,\\
g_2\left(E^x, E^y, \varepsilon\right)&=&c_2(\varepsilon,E^x/E^y)
\left(E^xE^y\right)^{-c/2}\,.
\end{eqnarray}
Here, $c_1$ is an integration constant while $c_2$ can be a function of
$\varepsilon$ and the ratio $E^x/E^y$ at most. If $c_2$ is not constant, we
have a version of inverse-triad corrections with a restriction on the triad
dependence analogous to what has been found in spherically symmetric models
\cite{JR}. (The two expressions $\epsilon$ and $E^x/E^y$ or functions of them
are the only combinations of triad components without density weight.) The
curvature dependence is not fully determined yet, but from (\ref{Sol6}) it
could only be of power-law form, already ruling out the usual choice of
periodic holonomy-modification functions. We will now show that only the
classical case $c=1$ of a linear dependence of $g_1$ on curvature components
is allowed.

We insert our solution for the correction function $g$
in (\ref{NGcond1}) and obtain
\begin{eqnarray}\label{Sol7}
  \frac{1}{E^x}\frac{\partial f}{\partial K_x} + \frac{1}{E^y}
  \frac{\partial f}{\partial K_y}=c_1
  \frac{\partial c_2}{\partial \varepsilon}\left[\sqrt{\frac{E^x}{E^y}}K_x
    +\sqrt{\frac{E^y}{E^x}}K_y\right]^c\,.
\end{eqnarray}
Doing the same in (\ref{Sol2}) yields
\begin{eqnarray}\label{Sol8}
\frac{1}{E^x}\frac{\partial f}{\partial K_x} - \frac{1}{E^y}
\frac{\partial f}{\partial K_y}&=&\frac{c c_1 c_2}{2\varepsilon}
\left[\sqrt{\frac{E^x}{E^y}}K_x +\sqrt{\frac{E^y}{E^x}}K_y\right]^c\\
&&\qquad\times
 \left[\frac{E^yK_y-E^xK_x}{E^xK_x+E^yK_y}-\frac{2}{c} \frac{E^x}{(E^y)^2}
  \frac{1}{c_2} \frac{\partial c_2}{\partial (E^x/E^y)}\right]\,. \nonumber
\end{eqnarray}
From these two relations, we identify the partial derivatives
\begin{eqnarray}
\frac{\partial f}{\partial K_x}&=&\frac{c_1
  E^x}{2}\left[\sqrt{\frac{E^x}{E^y}}K_x
  +\sqrt{\frac{E^y}{E^x}}K_y\right]^c\nonumber\\
&&\hspace{2cm}\times\left[\frac{\partial c_2}{\partial\varepsilon}+ \frac{c
    c_2}{2\varepsilon}
    \frac{E^yK_y-E^xK_x}{E^xK_x+E^yK_y} -\frac{c_2}{\varepsilon}
    \frac{E^x}{(E^y)^2} 
    \frac{1}{c_2} \frac{\partial c_2}{\partial (E^x/E^y)}
\right]\,,\label{Sol9}\\
\frac{\partial f}{\partial K_y}&=&\frac{c_1
  E^y}{2}\left[\sqrt{\frac{E^x}{E^y}}K_x
  +\sqrt{\frac{E^y}{E^x}}K_y\right]^c\nonumber\\
&&\hspace{2cm}\times\left[\frac{\partial c_2}{\partial \varepsilon}+ \frac{c
    c_2}{2\varepsilon}
    \frac{E^xK_x-E^yK_y}{E^xK_x+E^yK_y}+\frac{c_2}{\varepsilon}
    \frac{E^x}{(E^y)^2} 
    \frac{1}{c_2} \frac{\partial c_2}{\partial (E^x/E^y)}
\right]\,.\label{Sol10}
\end{eqnarray}
At this point, we still have a consistent system of equations. We can
calculate the left-hand side of (\ref{Sol3}) using the expressions above in
(\ref{Sol9}) and (\ref{Sol10}) and verify that it gives the same result as
the right-hand side of (\ref{Sol3}).

We now calculate the second-order mixed partial derivative by operating on
(\ref{Sol9}) with $\partial/\partial K_y$:
\begin{eqnarray}\label{NG1}
\frac{\partial^2 f}{\partial K_y\partial K_x}=&&\frac{c
  c_1}{2}\Bigg[\left\{\sqrt{\frac{E^x}{E^y}}K_x
  +\sqrt{\frac{E^y}{E^x}}K_y\right\}^{c-1}\sqrt{E^xE^y}
\left\{\frac{\partial c_2}{\partial \varepsilon} -\frac{1}{\epsilon}
  \frac{E^x}{(E^y)^2} \frac{\partial c_2}{\partial (E^x/E^y)}\right.\nonumber\\
&&  \hspace{7.5cm}\left.  +\frac{c c_2}{2\varepsilon}
  \left[\frac{-E^xK_x+E^yK_y}{E^xK_x+E^yK_y}\right]\right\}\nonumber\\
&& \hspace{2cm}+\frac{c_2 E^x}{\varepsilon}\left\{\sqrt{\frac{E^x}{E^y}}K_x
  +\sqrt{\frac{E^y}{E^x}}K_y\right\}^{c}
\left\{\frac{E^yE^xK_x}{(E^xK_x+E^yK_y)^2}\right\}\Bigg]\,.
\end{eqnarray}
We operate on (\ref{Sol10}) with $\partial/\partial K_x$ to obtain
\begin{eqnarray}\label{NG2}
\frac{\partial^2 f}{\partial K_x\partial K_y}=&&\frac{c
  c_1}{2}\Bigg[\left\{\sqrt{\frac{E^x}{E^y}}K_x
  +\sqrt{\frac{E^y}{E^x}}K_y\right\}^{c-1}\sqrt{E^xE^y}
\left\{\frac{\partial c_2}{\partial \varepsilon} +\frac{1}{\epsilon}
  \frac{E^x}{(E^y)^2} \frac{\partial c_2}{\partial (E^x/E^y)}\right.\nonumber\\
&&  \hspace{7.5cm}\left.+\frac{c c_2}{2\varepsilon}
  \left[\frac{E^xK_x-E^yK_y}{E^xK_x+E^yK_y}\right]\right\}\nonumber\\
&& \hspace{2cm}+\frac{c_2 E^y}{\varepsilon}\left\{\sqrt{\frac{E^x}{E^y}}K_x
  +\sqrt{\frac{E^y}{E^x}}K_y\right\}^{c}
\left\{\frac{E^xE^yK_y}{(E^xK_x+E^yK_y)^2}\right\}\Bigg]\,.
\end{eqnarray}
Requiring that these two quantities must be equal to each other results in one
fixed value of the constant, $c=1$. (Also, $\partial
c_2/\partial(E^x/E^y)=0$, so that $c_2$ depends only on $\epsilon$.)

Therefore, all modification functions that are consistent with anomaly freedom
have the classical dependence on curvature variables. It is impossible to
include holonomy modifications for these models with the parameterization
used. The only possibility left is to include holonomy-correction functions
modifying the dependence on all three variables, $K_x, K_y$ and
$\mathcal{A}$. It is not possible to factorize the holonomy function to give
separate point-wise correction functions and non-local ones. Moreover,
obstructions to this last possibility have been found in the related
expressions of spherically symmetric models \cite{HigherSpatial}.

It is instructive to look back at the spherically symmetric models and ask how
it is possible to introduce point-wise holonomy modifications in that
case. The answer lies in additional symmetries that ensure $E^x=E^y$. The
obstructions noted here can then be by-passed, but, as it appears, only as an
artifact of the more-symmetric nature of this model.  Quantizing a
symmetry-reduced model is different from symmetry-reducing a more general
quantum system, and accordingly we find additional obstructions to covariance
in our less-symmetric holonomy-modified models.

\subsubsection{Abelianization of normal deformations}
\label{s:Abel}

In the vacuum spherically symmetric model, an Abelianization of normal
hypersurface deformations has been found, making it easier to see consistent
modifications of the constraint \cite{LoopSchwarz}: One can use the
construction to eliminate most derivatives in the constraint, so that no
non-zero Poisson brackets remain with or without modified dependence on the
angular curvature component. If there is scalar matter, it is no longer
possible to eliminate as many spatial derivatives, and finding consistent
modifications is more complicated; in fact, so far only obstructions to
consistent modification have been seen \cite{SphSymmCov}. We now demonstrate
the analogous features for polarized Gowdy models: Classical Abelianization of
normal deformations is possible, but no consistent holonomy modification seems
to exist.

We write the constraints as
\begin{eqnarray}\label{clsclconstrnt}
&& H[N]=-\frac{1}{8\pi G}\int \text{d}\theta\,\,
N\Bigg[K_xK_y\varepsilon^{-1/2}\sqrt{E^xE^y} +
\varepsilon^{1/2}\left(\sqrt{\frac{E^x}{E^y}}K_x+
\sqrt{\frac{E^y}{E^x}}K_y\right)\mathcal{A}\nonumber\\
&&\hspace{3cm}+\frac{1}{4\sqrt{\varepsilon
    E^xE^y}}\left(\left[\varepsilon^{\prime}\right]^2
  -\left[\varepsilon\left(\ln{\frac{E^y}{E^x}}\right)^{\prime}\right]^2\right)
-\left(\frac{\varepsilon^{1/2}\varepsilon^{\prime}}{\sqrt{E^xE^y}}
\right)^{\prime}\Bigg]\\
&& D[N^\theta]=\frac{1}{8\pi G}\int \text{d}\theta\,\,
N^\theta\left[K_x^{\prime}E^x+ K_y^{\prime}E^y
  -\varepsilon^{\prime}\mathcal{A}\right]\,.
\end{eqnarray}
They can be combined to the total constraint
\begin{equation}\label{totham}
  \text{H}_{\text{T}}[N,N^{\theta}]=\frac{1}{\kappa}\int\text{d}\theta\,
  \left[-N(\theta)\mathcal{H}(\theta)
    +N^\theta(\theta)\mathcal{D}(\theta)\right]\,,
\end{equation}
where $\mathcal{H}$ and $\mathcal{D}$ are the unsmeared local versions of the
gravitational constraints (\ref{clsclconstrnt}).

We keep $\mathcal{D}$ as a constraint but replace ${\cal H}$  by the linear
combination
\begin{equation}
 {\cal C}= \frac{\epsilon'}{\sqrt{E^xE^y}} \mathcal{H}+ \sqrt{\epsilon}
 \left(\frac{K_x}{E^y}+\frac{K_y}{E^x}\right)\mathcal{D}\,,
\end{equation}
smeared to a new constraint
\begin{eqnarray}\label{newHam}
&& C[L]=-\frac{1}{8\pi G}\int \text{d}\theta\,\,
L\Bigg[K_xK_y\varepsilon^{-1/2}\varepsilon^{\prime} +
\varepsilon^{1/2}\left(K_xK_y^{\prime}+K_yK_x^{\prime}+
\left[\frac{E^x}{E^y}\right]K_xK_x^{\prime}+
  \left[\frac{E^y}{E^x}\right]K_yK_y^{\prime}\right)\nonumber\\
&&\hspace{3cm}+\frac{\varepsilon^{\prime}}{4\sqrt{\varepsilon}
  E^xE^y}\left(\left[\varepsilon^{\prime}\right]^2
  -\left[\varepsilon\left(\ln{\frac{E^y}{E^x}}\right)^{\prime}\right]^2\right)
-\left(\frac{\varepsilon^{1/2}\varepsilon^{\prime}}{\sqrt{E^xE^y}}
\right)^{\prime}\Bigg]\,.
\end{eqnarray}
As in Abelianizations of normal deformations in spherically symmetric models
\cite{LoopSchwarz,LoopSchwarz2}, an important feature of the new constraint is
that the inhomogeneous curvature component, here $\mathcal{A}$, has been
eliminated.

Computing the brackets of constraints $(C[L],D[N^{\theta}])$, it is clear that
the $\{D,D\}$-bracket has the original form. Also the $\{C,D\}$-bracket
has the same form as the original $\{H,D\}$-bracket because $\mathcal{C}$ has
the same spatial density weight as $\mathcal{H}$. The $\{C,C\}$-bracket must
be computed explicitly, and turns out to be zero as shown in
App.~\ref{a:Abel}. See also \cite{GowdyAbel} for a related
  result. The set of brackets of the constraints takes the form
\begin{eqnarray}
\left\{D[N^\theta],D[M^\theta]\right\}&=&D[\mathcal{L}_{N^\theta}M^\theta]\nonumber\\
\left\{C[L],D[M^\theta]\right\}&=&-C[\mathcal{L}_{M^\theta}L]\nonumber\\
\left\{C[L_1],C[L_2]\right\}&=&0\,.
\end{eqnarray}
As in spherically symmetric models, one cannot consistently modify the
curvature dependence of the constraints without destroying properties relevant
for closure of the brackets.

\subsection{Relation to hybrid models}

A Gowdy system has been proposed and analyzed in the context of loop quantum
gravity in a hybrid version \cite{Hybrid,Hybrid2,Hybrid3}: There is a
homogeneous background with modifications suggested by loop quantum cosmology,
coupled to inhomogeneous Gowdy modes quantized in the standard way on a Fock
space. Concrete realizations make use of gauge fixings of space-time
transformations, but nevertheless the framework should be expected to be
covariant: It is an example of a covariant quantum-field theory (the
Fock-represented Gowdy modes) on a Riemannian background (the loop-modified
homogeneous model). Since quantum-field theory has an established covariant
formulation on any curved background, not just on those satisfying Einstein's
equation, there is no reason why covariance should be broken in hybrid models,
interpreted as systems of quantum fields on a background. Indeed, different
choices of gauge fixings have been shown to lead to compatible results
\cite{HybridMuk}.

However, going beyond the background setting is more difficult. (See
\cite{SphSymmCov} for a detailed discussion of the difference between
background treatments and background-independent models in the context of
modified or quantized canonical theories.) To do so, one would have to show
that the modified background can be part of a covariant inhomogeneous model of
Gowdy type. Our no-go results show that this condition is difficult to
achieve. It therefore seems unlikely that hybrid models can be reductions of a
covariant background-independent system with the same symmetries (leaving
aside the much harder question of a reduction from a covariant full
theory). Such an extension would be important not just on conceptual grounds,
but also for a uniform treatment of modifications: In hybrid models, the
background dynamics is modified by loop effects (holonomies), but
inhomogeneous mode equations have no such modifications (except indirect ones
via background variables in their coefficients). When holonomy effects are
significant for the background dynamics (near a ``bounce'' at large
curvature), they should be expected to contribute to the dynamics of
inhomogeneities as well. (Interestingly, numerical investigations in hybrid
models have revealed instabilities \cite{InhomThroughBounce} reminiscent of
some effects related to signature change \cite{Action,SigChange,SigImpl}, an
apparently generic consequence of consistent holonomy modifications of
inhomogeneous gravitational equations \cite{JR,ScalarHolInv,DeformedCosmo}.)
Consistently including these terms in inhomogeneous equations requires a
covariant Gowdy model with holonomy modifications, which has failed to
materialize in our attempts shown here. Using our partial no-go results,
several non-trivial modifications would be required to ensure covariance,
which go well beyond those included in our already rather general functions
$f$, $g_1$ and $g_2$.

\section{Conclusions}

In this paper, we have continued the discussion of covariance in
holonomy-modified models with local degrees of freedom, started in
\cite{SphSymmCov} for spherically symmetric models with matter. Also here,
partial no-go results but no consistent covariant versions have been
found. One cannot draw final conclusions from partial no-go results, but they
do show that holonomy modifications in inhomogeneous models cannot be as
simple as they had been anticipated in homogeneous models. In the models
studied here and in \cite{SphSymmCov}, covariance is therefore shown to be a
restrictive criterion, capable of limiting the quantization choices that exist
without the condition (as emphasized for instance in
\cite{AlexAmbig}). However, at present it is not clear whether holonomy
modifications in models with local degrees of freedom can lead to covariant
theories at all. Further study into this question and the related problem of
anomalies in canonical quantum gravity is needed before the effects proposed
in homogeneous models can be considered generic. As in \cite{SphSymmCov}, it
is encouraging that the analysis of Poisson brackets of modified constraints
leads to the same result as attempts to Abelianize the generators of normal
hypersurface deformations, which has been shown in Sec.~\ref{s:Abel} to be
possible for classical polarized Gowdy models, but not for the proposed
modified ones.

Even though the modifications used here did not lead to fully covariant
models, we were able to confirm certain structural properties of constraint
brackets in the extension to Gowdy systems. If conditions for anomaly freedom
are only partially solved so as to allow for non-trivial modifications, as
analyzed in the first part of Sec.~\ref{s:Structure}, the multiplier of the
diffeomorphism constraint in the bracket of two modified Hamiltonian
constraints receives a factor (\ref{beta}) which is negative around a local
maximum of the holonomy-modification function. The presence of anomalies means
that this statement cannot be a physical one as long as no consistent set of
modified constraints has been found. Nevertheless, the dependence of
modification functions on two independent variables makes the behavior of
local maxima less trivial than in the case of spherically symmetric
models. The fact that the same formal behavior is found is an indication that
the sign of the multiplier around local maxima may be generic, as would be the
consequence of signature change.

\section*{Acknowledgements}

We are grateful to Guillermo Mena Marug\'an for discussions and comments on a
draft of this paper.  This work was supported in part by NSF grant
PHY-1307408.

\begin{appendix}

\section{Abelian constraints}
\label{a:Abel}

In order to confirm the Abelian nature of (\ref{newHam}), we list all
non-trivial terms in the $\{C,C\}$-bracket, split in different types according
to the ``kinetic'' terms in $C$ (that is, those containing extrinsic curvature
components). The non-zero Poisson brackets from terms of the form
$\{K_i,E^i\}$ cancel out by antisymmetry. The only remaining non-zero terms
come from the $\{K_i,E^{i\prime}\}$, $\{K_i^{\prime},E^{i}\}$ and
$\{K_i^{\prime},E^{i\prime}\}$-types of brackets, where $i$ can be either $x$
or $y$.

Terms of the first kind are:
\begin{eqnarray}
&&\frac{\varepsilon(\varepsilon^{\prime})^2K_xE^{y\prime}}{2E^x(E^y)^3}+
\frac{\varepsilon(\varepsilon^{\prime})^2K_yE^{x\prime}}{2(E^x)^3E^y}-
\frac{\varepsilon(\varepsilon^{\prime})^2K_yE^{y\prime}}{2(E^x)^2(E^y)^2} -
\frac{\varepsilon(\varepsilon^{\prime})^2K_xE^{x\prime}}{2(E^x)^2(E^y)^2} -
\frac{(\varepsilon^{\prime})^3K_y}{2(E^x)^2E^y}\nonumber\\
&& - \frac{(\varepsilon^{\prime})^3K_x}{2E^x(E^y)^2}+
\frac{(\varepsilon)^2\varepsilon^{\prime}K_y^{\prime}E^{x\prime}}{2(E^x)^3E^y}
-\frac{(\varepsilon)^2\varepsilon^{\prime}K_y^{\prime}E^{y\prime}}{2(E^x)^2(E^y)^2}
-\frac{\varepsilon(\varepsilon^{\prime})^2K_y^{\prime}}{2(E^x)^2E^y} +
\frac{(\varepsilon)^2\varepsilon^{\prime}K_x^{\prime}E^{y\prime}}{2E^x(E^y)^3}
\nonumber\\
&&-\frac{(\varepsilon)^2\varepsilon^{\prime}K_x^{\prime}E^{x\prime}}{2(E^x)^2(E^y)^2}
-\frac{\varepsilon(\varepsilon^{\prime})^2K_x^{\prime}}{2E^x(E^y)^2}
+\frac{(\varepsilon)^2\varepsilon^{\prime}K_x^{\prime}E^{x\prime}}{2(E^x)^2(E^y)^2}
-\frac{(\varepsilon)^2\varepsilon^{\prime}K_x^{\prime}E^{y\prime}}{2E^x(E^y)^3}
-\frac{\varepsilon(\varepsilon^{\prime})^2K_x^{\prime}}{2E^x(E^y)^2}\nonumber\\
&&+\frac{(\varepsilon)^2\varepsilon^{\prime}K_y^{\prime}E^{y\prime}}{2(E^x)^2(E^y)^2}
-\frac{(\varepsilon)^2\varepsilon^{\prime}K_y^{\prime}E^{x\prime}}{2(E^x)^3E^y}
-\frac{\varepsilon(\varepsilon^{\prime})^2K_y^{\prime}}{2(E^x)^2E^y}\,.
\end{eqnarray}

Terms of the second kind are:
\begin{eqnarray}
&&-\frac{\varepsilon K_x^2K_x^{\prime}E^x}{(E^y)^2}+ \frac{\varepsilon
  K_xK_yK_y^{\prime}}{E^x}+ \frac{(\varepsilon^{\prime})^3K_x}{4E^x(E^y)^2}+
\frac{3(\varepsilon)^2\varepsilon^{\prime}(E^{y\prime})^2K_x}{4E^x(E^y)^4}
+\frac{(\varepsilon)^2\varepsilon^{\prime}(E^{x\prime})^2K_x}{4(E^x)^3(E^y)^2}
\nonumber\\
&&-\frac{(\varepsilon)^2\varepsilon^{\prime}E^{x\prime}E^{y\prime}K_x}{(E^x)^2(E^y)^3}
-\frac{\varepsilon(\varepsilon^{\prime})^2E^{x\prime}K_x}{2(E^x)^2(E^y)^2}
-\frac{\varepsilon(\varepsilon^{\prime})^2E^{y\prime}K_x}{E^x(E^y)^3} +
\frac{\varepsilon\varepsilon^{\prime}\varepsilon^{\prime\prime}K_x}{E^x(E^y)^2}
\nonumber\\
&&+\frac{\varepsilon K_yK_xK_x^{\prime}}{E^y}- \frac{\varepsilon
  E^{y}K_y^2K_y^{\prime}}{(E^x)^2}
+\frac{(\varepsilon^{\prime})^3K_y}{4(E^x)^2E^y}
+\frac{(\varepsilon)^2\varepsilon^{\prime}(E^{y\prime})^2K_y}{4(E^x)^2(E^y)^3}
+\frac{3(\varepsilon)^2\varepsilon^{\prime}(E^{x\prime})^2K_y}{4(E^x)^4E^y}
\nonumber\\
&&-\frac{(\varepsilon)^2\varepsilon^{\prime}E^{x\prime}E^{y\prime}K_y}{(E^x)^3(E^y)^2}
-\frac{\varepsilon(\varepsilon^{\prime})^2E^{y\prime}K_y}{2(E^x)^2(E^y)^2}
-\frac{\varepsilon(\varepsilon^{\prime})^2E^{x\prime}K_y}{(E^x)^3E^y} +
\frac{\varepsilon\varepsilon^{\prime}\varepsilon^{\prime\prime}K_y}{(E^x)^2E^y}
\nonumber\\
&&+\frac{\varepsilon K_x^2K_x^{\prime}E^x}{(E^y)^2}- \frac{\varepsilon
  K_xK_yK_y^{\prime}}{E^x}+ \frac{(\varepsilon^{\prime})^3K_x}{4E^x(E^y)^2}+
\frac{3(\varepsilon)^2\varepsilon^{\prime}(E^{x\prime})^2K_x}{4(E^x)^3(E^y)^2}
+\frac{(\varepsilon)^2\varepsilon^{\prime}(E^{y\prime})^2K_x}{4E^x(E^y)^4}
\nonumber\\
&&-\frac{(\varepsilon)^2\varepsilon^{\prime}E^{x\prime}E^{y\prime}K_x}{(E^x)^2(E^y)^3}
-\frac{\varepsilon(\varepsilon^{\prime})^2E^{x\prime}K_x}{(E^x)^2(E^y)^2}
-\frac{\varepsilon(\varepsilon^{\prime})^2E^{y\prime}K_x}{2E^x(E^y)^3} +
\frac{\varepsilon\varepsilon^{\prime}\varepsilon^{\prime\prime}K_x}{E^x(E^y)^2}
\nonumber\\
&&-\frac{\varepsilon K_yK_xK_x^{\prime}}{E^y}+ \frac{\varepsilon
  E^{y}K_y^2K_y^{\prime}}{(E^x)^2}
+\frac{(\varepsilon^{\prime})^3K_y}{4(E^x)^2E^y}
+\frac{3(\varepsilon)^2\varepsilon^{\prime}(E^{y\prime})^2K_y}{4(E^x)^2(E^y)^3}
+\frac{(\varepsilon)^2\varepsilon^{\prime}(E^{x\prime})^2K_y}{4(E^x)^4E^y}
\nonumber\\
&&-\frac{(\varepsilon)^2\varepsilon^{\prime}E^{x\prime}E^{y\prime}K_y}{(E^x)^3(E^y)^2}
-\frac{\varepsilon(\varepsilon^{\prime})^2E^{y\prime}K_y}{(E^x)^2(E^y)^2}
-\frac{\varepsilon(\varepsilon^{\prime})^2E^{x\prime}K_y}{2(E^x)^3E^y} +
\frac{\varepsilon\varepsilon^{\prime}\varepsilon^{\prime\prime}K_y}{(E^x)^2E^y}\,.
\end{eqnarray}

And finally, the most complicated terms come from brackets of the form
$\{K_i^{\prime},E^{i\prime}\}$. Since there are many terms of this form, we
first list those from the contributions proportional to $K_xK_y^{\prime}$ and
$K_yK_x^{\prime}$:
\begin{eqnarray}
&&-\frac{\varepsilon(\varepsilon^{\prime})^2E^{y\prime}K_x}{2E^x(E^y)^3}
-\frac{(\varepsilon)^2\varepsilon^{\prime}E^{y\prime}K_x^{\prime}}{2E^x(E^y)^3}
+\frac{(\varepsilon)^2\varepsilon^{\prime\prime}E^{y\prime}K_x}{2E^x(E^y)^3}
+\frac{(\varepsilon)^2\varepsilon^{\prime}E^{y\prime\prime}K_x}{2E^x(E^y)^3}
+\frac{(\varepsilon)^2\varepsilon^{\prime}E^{x\prime}E^{y\prime}K_x}{2(E^x)^2(E^y)^3}
\nonumber\\
&&-\frac{3(\varepsilon)^2\varepsilon^{\prime}(E^{y\prime})^2K_x}{2E^x(E^y)^4}
+\frac{(\varepsilon)^2\varepsilon^{\prime}E^{x\prime}K_x^{\prime}}{2(E^x)^2(E^y)^2}
-\frac{(\varepsilon)^2\varepsilon^{\prime\prime}E^{x\prime}K_x}{2(E^x)^2(E^y)^2}
-\frac{(\varepsilon)^2\varepsilon^{\prime}E^{x\prime\prime}K_x}{2(E^x)^2(E^y)^2}
+\frac{(\varepsilon)^2\varepsilon^{\prime}(E^{x\prime})^2K_x}{(E^x)^3(E^y)^2}
\nonumber\\
&&+\frac{\varepsilon(\varepsilon^{\prime})^2K_x^{\prime}}{2E^x(E^y)^2}
-\frac{\varepsilon\varepsilon^{\prime}\varepsilon^{\prime\prime}K_x}{E^x(E^y)^2}
\nonumber\\
&&-\frac{\varepsilon(\varepsilon^{\prime})^2E^{x\prime}K_y}{2E^y(E^x)^3}
-\frac{(\varepsilon)^2\varepsilon^{\prime}E^{x\prime}K_y^{\prime}}{2E^y(E^x)^3}
+\frac{(\varepsilon)^2\varepsilon^{\prime\prime}E^{x\prime}K_y}{2E^y(E^x)^3}
+\frac{(\varepsilon)^2\varepsilon^{\prime}E^{x\prime\prime}K_y}{2E^y(E^x)^3}
+\frac{(\varepsilon)^2\varepsilon^{\prime}E^{x\prime}E^{y\prime}K_y}{2(E^y)^2(E^x)^3}
\nonumber\\
&&-\frac{3(\varepsilon)^2\varepsilon^{\prime}(E^{x\prime})^2K_y}{2E^y(E^x)^4}
+\frac{(\varepsilon)^2\varepsilon^{\prime}E^{y\prime}K_y^{\prime}}{2(E^y)^2(E^x)^2}
-\frac{(\varepsilon)^2\varepsilon^{\prime\prime}E^{y\prime}K_y}{2(E^x)^2(E^y)^2}
-\frac{(\varepsilon)^2\varepsilon^{\prime}E^{y\prime\prime}K_y}{2(E^x)^2(E^y)^2}
+\frac{(\varepsilon)^2\varepsilon^{\prime}(E^{y\prime})^2K_y}{(E^y)^3(E^x)^2}
\nonumber\\
&&+\frac{\varepsilon(\varepsilon^{\prime})^2K_y^{\prime}}{2E^y(E^x)^2}
-\frac{\varepsilon\varepsilon^{\prime}\varepsilon^{\prime\prime}K_y}{E^y(E^x)^2}\,.
\end{eqnarray}

The other two kinetic terms proportional to $K_xK_x^{\prime}$ and
$K_yK_y^{\prime}$ also give contributions via the
$\{K_i^{\prime},E^{i\prime}\}$-bracket:
\begin{eqnarray}
&&\frac{2\varepsilon(\varepsilon^{\prime})^2E^{x\prime}K_x}{(E^x)^2(E^y)^2}
-\frac{2(\varepsilon)^2\varepsilon^{\prime}(E^{x\prime})^2K_x}{(E^x)^3(E^y)^2}
+\frac{3(\varepsilon)^2\varepsilon^{\prime}E^{x\prime}E^{y\prime}K_x}{2(E^x)^2(E^y)^3}
-\frac{(\varepsilon)^2\varepsilon^{\prime}E^{x\prime}K_x^{\prime}}{2(E^x)^2(E^y)^2}
+\frac{(\varepsilon)^2\varepsilon^{\prime\prime}E^{x\prime}K_x}{2(E^x)^2(E^y)^2}
\nonumber\\
&&+\frac{(\varepsilon)^2\varepsilon^{\prime}E^{x\prime\prime}K_x}{2(E^x)^2(E^y)^2}
-\frac{\varepsilon(\varepsilon^{\prime})^2E^{y\prime}K_x}{2E^x(E^y)^3}
+\frac{(\varepsilon)^2\varepsilon^{\prime}(E^{y\prime})^2K_x}{2E^x(E^y)^4}
+\frac{(\varepsilon)^2\varepsilon^{\prime}E^{y\prime}K_x^{\prime}}{2E^x(E^y)^3}
-\frac{(\varepsilon)^2\varepsilon^{\prime\prime}E^{y\prime}K_x}{2E^x(E^y)^3}
\nonumber\\
&&-\frac{(\varepsilon)^2\varepsilon^{\prime}E^{y\prime\prime}K_x}{2E^x(E^y)^3}
+\frac{\varepsilon(\varepsilon^{\prime})^2K_x^{\prime}}{2E^x(E^y)^2}
-\frac{\varepsilon\varepsilon^{\prime}\varepsilon^{\prime\prime}K_x}{E^x(E^y)^2}
\nonumber\\
&&\frac{2\varepsilon(\varepsilon^{\prime})^2E^{y\prime}K_y}{(E^x)^2(E^y)^2}
-\frac{2(\varepsilon)^2\varepsilon^{\prime}(E^{y\prime})^2K_y}{(E^y)^3(E^x)^2}
+\frac{3(\varepsilon)^2\varepsilon^{\prime}E^{x\prime}E^{y\prime}K_y}{2(E^y)^2(E^x)^3}
-\frac{(\varepsilon)^2\varepsilon^{\prime}E^{y\prime}K_y^{\prime}}{2(E^x)^2(E^y)^2}
+\frac{(\varepsilon)^2\varepsilon^{\prime\prime}E^{y\prime}K_y}{2(E^x)^2(E^y)^2}
\nonumber\\
&&+\frac{(\varepsilon)^2\varepsilon^{\prime}E^{y\prime\prime}K_y}{2(E^x)^2(E^y)^2}
-\frac{\varepsilon(\varepsilon^{\prime})^2E^{x\prime}K_y}{2E^y(E^x)^3}
+\frac{(\varepsilon)^2\varepsilon^{\prime}(E^{x\prime})^2K_y}{2E^y(E^x)^4}
+\frac{(\varepsilon)^2\varepsilon^{\prime}E^{x\prime}K_y^{\prime}}{2E^y(E^x)^3}
-\frac{(\varepsilon)^2\varepsilon^{\prime\prime}E^{x\prime}K_y}{2E^y(E^x)^3}
\nonumber\\
&&-\frac{(\varepsilon)^2\varepsilon^{\prime}E^{x\prime\prime}K_y}{2E^y(E^x)^3}
+\frac{\varepsilon(\varepsilon^{\prime})^2K_y^{\prime}}{2E^y(E^x)^2}
-\frac{\varepsilon\varepsilon^{\prime}\varepsilon^{\prime\prime}K_y}{E^y(E^x)^2}\,.
\end{eqnarray}

These are all non-zero terms, and in spite of their large number it is
straightforward to observe that they all cancel one another when combined.

\end{appendix}


\begin{thebibliography}{10}

\bibitem{SphSymmCov}
M.\ Bojowald, S.\ Brahma, and J.~D.\ Reyes,
\newblock Covariance in models of loop quantum gravity: Spherical symmetry,
  [arXiv:1507.00329]

\bibitem{Hybrid}
M.\ Mart{\'\i}n-Benito, L.~J.\ Garay, and G.~A.\ Mena~Marug\'an,
\newblock Hybrid Quantum Gowdy Cosmology: Combining Loop and Fock
  Quantizations,
\newblock {\em Phys.\ Rev.\ D} 78 (2008) 083516, [arXiv:0804.1098]

\bibitem{Hybrid2}
L.~J.\ Garay, M.\ Mart\'{\i}n-Benito, and G.~A.\ Mena~Marug\'an,
\newblock Inhomogeneous Loop Quantum Cosmology: Hybrid Quantization of the
  Gowdy Model,
\newblock {\em Phys.\ Rev.\ D} 82 (2010) 044048, [arXiv:1005.5654]

\bibitem{Hybrid3}
M.\ Mart{\'\i}n-Benito, D.\ Mart\'{\i}n-de Blas, and G.~A.\ Mena~Marug\'an,
\newblock Matter in inhomogeneous loop quantum cosmology: the Gowdy $T^3$
  model,
\newblock {\em Phys.\ Rev.\ D} 83 (2011) 084050, [arXiv:1012.2324]

\bibitem{MisnerGowdy}
C.~W.\ Misner,
\newblock A Minisuperspace Example: The Gowdy $T^3$ Cosmology,
\newblock {\em Phys.\ Rev.\ D} 8 (1973) 3271--3285

\bibitem{GowdyQuadratic}
B.~K.\ Berger,
\newblock Quantum Cosmology: Exact Solution For The Gowdy $T^3$ Model,
\newblock {\em Phys.\ Rev.\ D} 11 (1975) 2770--2780

\bibitem{BergerGowdy}
B.~K.\ Berger,
\newblock Quantum Effects In The Gowdy $T^3$ Cosmology,
\newblock {\em Ann.\ Phys.} 156 (1984) 155--193

\bibitem{HusainGowdySing}
V.\ Husain,
\newblock Quantum Effects on the Singularity of the Gowdy Cosmology,
\newblock {\em Class.\ Quantum Grav.} 4 (1987) 1587--1591

\bibitem{HusainSmolin}
V.\ Husain and L.\ Smolin,
\newblock Exactly Solvable Quantum Cosmologies From Two Killing Field
  Reductions Of General Relativity,
\newblock {\em Nucl.\ Phys.\ B} 327 (1989) 205

\bibitem{GowdyQuant}
G.\ Mena~Marug\'an,
\newblock Canonical quantization of the Gowdy model,
\newblock {\em Phys.\ Rev.\ D} 56 (1997) 908--919, [gr-qc/9704041]

\bibitem{Probe2}
M.\ Pierri,
\newblock Probing quantum general relativity through exactly soluble
  midi-superspaces. II: Polarized Gowdy models,
\newblock {\em Int.\ J.\ Mod.\ Phys.\ D} 11 (2002) 135, [gr-qc/0101013]

\bibitem{GowdyTime}
A.\ Corichi, J.\ Cortez, and H.\ Quevedo,
\newblock On time evolution in quantum Gowdy $T^3$ models,
\newblock {\em Phys.\ Rev.\ D} 67 (2003) 087502

\bibitem{GowdyDyn}
C.~G.\ Torre,
\newblock Quantum dynamics of the polarized Gowdy $T^3$ model,
\newblock {\em Phys.\ Rev.\ D} 66 (2002) 084017, [gr-qc/0206083]

\bibitem{GowdyDynSchroed}
C.~G.\ Torre,
\newblock Schr\"odinger representation for the polarized Gowdy model,
  [gr-qc/0607084]

\bibitem{LivRev}
M.\ Bojowald,
\newblock Loop Quantum Cosmology,
\newblock {\em Living Rev.\ Relativity} 11 (2008) 4, [gr-qc/0601085],
\newblock {\tt http://www.livingreviews.org/lrr-2008-4}

\bibitem{ROPP}
M.\ Bojowald,
\newblock Quantum cosmology: a review,
\newblock {\em Rep.\ Prog.\ Phys.} 78 (2015) 023901, [arXiv:1501.04899]

\bibitem{Gowdy}
R.~H.\ Gowdy,
\newblock Vacuum spacetimes with two-parameter spacelike isometry groups and
  compact invariant hypersurfaces: Topologies and boundary conditions,
\newblock {\em Ann.\ Phys.} 83 (1974) 203--241

\bibitem{DiracHamGR}
P.~A.~M.\ Dirac,
\newblock The theory of gravitation in Hamiltonian form,
\newblock {\em Proc.\ Roy.\ Soc.\ A} 246 (1958) 333--343

\bibitem{ADM}
R.\ Arnowitt, S.\ Deser, and C.~W.\ Misner,
\newblock The Dynamics of General Relativity, In L.\ Witten, editor, {\em
  Gravitation: An Introduction to Current Research},
\newblock Wiley, New York, 1962,
\newblock Reprinted in \cite{ADMRe}

\bibitem{LoopSchwarz2}
R.\ Gambini and J.\ Pullin,
\newblock Hawking radiation from a spherical loop quantum gravity black hole,
  [arXiv:1312.3595]

\bibitem{KucharHypI}
K.~V.\ Kucha\v{r},
\newblock Geometry of hypersurfaces. I,
\newblock {\em J.\ Math.\ Phys.} 17 (1976) 777--791

\bibitem{Energy}
M.\ Bojowald, G.\ Hossain, M.\ Kagan, and C.\ Tomlin,
\newblock Quantum matter in quantum space-time,
\newblock {\em Quantum Matter} 2 (2013) 436--443, [arXiv:1302.5695]

\bibitem{SphSymm}
M.\ Bojowald,
\newblock Spherically Symmetric Quantum Geometry: States and Basic Operators,
\newblock {\em Class.\ Quantum Grav.} 21 (2004) 3733--3753, [gr-qc/0407017]

\bibitem{SphSymmHam}
B.~K.\ Berger, D.~M.\ Chitre, V.~E.\ Moncrief, and Y.\ Nutku,
\newblock Hamiltonian formulation of spherically symmetric gravitational
  fields,
\newblock {\em Phys.\ Rev.\ D} 5 (1972) 2467--2470

\bibitem{JR}
J.~D.\ Reyes,
\newblock {\em Spherically Symmetric Loop Quantum Gravity: Connections to
  2-Dimensional Models and Applications to Gravitational Collapse},
\newblock PhD thesis, The Pennsylvania State University, 2009

\bibitem{HigherSpatial}
M.\ Bojowald, G.~M.\ Paily, and J.~D.\ Reyes,
\newblock Discreteness corrections and higher spatial derivatives in effective
  canonical quantum gravity,
\newblock {\em Phys.\ Rev.\ D} 90 (2014) 025025, [arXiv:1402.5130]

\bibitem{Action}
M.\ Bojowald and G.~M.\ Paily,
\newblock Deformed General Relativity and Effective Actions from Loop Quantum
  Gravity,
\newblock {\em Phys.\ Rev.\ D} 86 (2012) 104018, [arXiv:1112.1899]

\bibitem{SigChange}
J.\ Mielczarek,
\newblock Signature change in loop quantum cosmology,
\newblock {\em Springer Proc.\ Phys.} 157 (2014) 555, [arXiv:1207.4657]

\bibitem{SigImpl}
M.\ Bojowald and J.\ Mielczarek,
\newblock Some implications of signature-change in cosmological models of loop
  quantum gravity, [arXiv:1503.09154]

\bibitem{EinsteinRosenAsh}
K.\ Banerjee and G.\ Date,
\newblock Loop quantization of polarized Gowdy model on $T^3$: Classical
  theory,
\newblock {\em Class.\ Quantum Grav.} 25 (2008) 105014, [arXiv:0712.0683]

\bibitem{ScalarHolInv}
T.\ Cailleteau, L.\ Linsefors, and A.\ Barrau,
\newblock Anomaly-free perturbations with inverse-volume and holonomy
  corrections in Loop Quantum Cosmology,
\newblock {\em Class.\ Quantum Grav.} 31 (2014) 125011, [arXiv:1307.5238]

\bibitem{DeformedCosmo}
A.\ Barrau, M.\ Bojowald, G.\ Calcagni, J.\ Grain, and M.\ Kagan,
\newblock Anomaly-free cosmological perturbations in effective canonical
  quantum gravity, [arXiv:1404.1018]

\bibitem{Loss}
M.\ Bojowald,
\newblock Information loss, made worse by quantum gravity,
\newblock {\em Front.\ Phys.} 3 (2015) 33, [arXiv:1409.3157]

\bibitem{LoopSchwarz}
R.\ Gambini and J.\ Pullin,
\newblock Loop quantization of the Schwarzschild black hole,
\newblock {\em Phys.\ Rev.\ Lett.} 110 (2013) 211301, [arXiv:1302.5265]

\bibitem{GowdyAbel}
D.\ Mart\'{\i}n-de Blas, J.\ Olmedo, and T.\ Pawlowski,
\newblock Loop quantization of the Gowdy model with local rotational symmetry,
  [in preparation]

\bibitem{HybridMuk}
L.\ Castell\'o~Gomar, M.\ Fern\'andez-M\'endez, G.~A.\ Mena~Marug\'an, and J.\
  Olmedo,
\newblock Cosmological perturbations in Hybrid Loop Quantum Cosmology:
  Mukhanov--Sasaki variables,
\newblock {\em Phys.\ Rev.\ D} 90 (2014) 064015, [arXiv:1407.0998]

\bibitem{InhomThroughBounce}
D.\ Brizuela, G.~A.\ Mena~Marug\'an, and T.\ Pawlowski,
\newblock Big Bounce and inhomogeneities,
\newblock {\em Class.\ Quantum Grav.} 27 (2010) 052001, [arXiv:0902.0697]

\bibitem{AlexAmbig}
A.\ Perez,
\newblock On the regularization ambiguities in loop quantum gravity,
\newblock {\em Phys.\ Rev.\ D} 73 (2006) 044007, [gr-qc/0509118]

\bibitem{ADMRe}
R.\ Arnowitt, S.\ Deser, and C.~W.\ Misner,
\newblock The Dynamics of General Relativity,
\newblock {\em Gen.\ Rel.\ Grav.} 40 (2008) 1997--2027

\end{thebibliography}

\end{document}